\documentclass[twocolumn,
aps,
prx,
showpacs,
amsmath,
amssymb,
floatfix,
longbibliography,
superscriptaddress,
eprint]
{revtex4-1}

\usepackage{subfigure}
\usepackage{amsfonts}

\usepackage{xcolor}
\definecolor{darkblue}{RGB}{0,0,150}
\definecolor{nightblue}{RGB}{0,0,100}

\usepackage{graphicx,mathtools,bm,bbm}
\usepackage{tikz}
\graphicspath{{./}}
\usepackage{booktabs}
\usepackage{MnSymbol}
\usepackage[
colorlinks,
citecolor=darkblue,
linkcolor=darkblue,
urlcolor=nightblue]{hyperref}
\usepackage{xurl}

\usepackage[english]{babel}
\usepackage[babel,kerning=true,spacing=true]{microtype}
\usepackage[utf8]{inputenc}

\usepackage{soul}

\begin{document}

\title{Color code off-the-hook: avoiding hook errors with a single auxiliary per plaquette}

\author{Gilad Kishony}
\affiliation{Classiq Technologies. 3 Daniel Frisch Street, Tel Aviv-Yafo, 6473104, Israel.}
\author{Austin Fowler}

\begin{abstract}

Syndrome extraction in the planar color code is complicated by high weight stabilizers and hook errors that can reduce the circuit-level distance.
With a single auxiliary qubit per plaquette, any spatially uniform circuit halves the circuit-level distance.
We propose a single-auxiliary syndrome extraction circuit with color-dependent gate schedules that avoids all malign hook errors in the bulk, thereby preserving the full circuit-level distance.
The circuit has minimal depth: all stabilizers of the same Pauli type are measured in parallel in six time steps. 
Furthermore, this schedule can be readily applied to the XYZ color code circuit, yielding an improved temporal distance.
We find that at the boundary, no single hook error alone reduces the distance; instead, only certain combinations of hook errors do, which we call fractional hook errors.
We demonstrate through Monte Carlo simulations over a range of circuit-level noise models and physical error rates that our circuit outperforms the previous state of the art.

\end{abstract}

\maketitle

\section{Introduction}
\label{sec:intro}

The color code~\cite{bombin2006color} is a topological quantum error correcting code that, like the surface code~\cite{dennis2002topological, fowler2012surface}, can be implemented on planar hardware.
It is an attractive candidate for fault-tolerant quantum computation because Clifford gates can be implemented transversally~\cite{bombin2006color}, it requires fewer physical qubits than the surface code for the same code distance~\cite{Thomsen_2024}, and it admits efficient magic state cultivation~\cite{gidney2024magicstatecultivationgrowing}.
However, syndrome extraction in the color code is more difficult than in the surface code for two main reasons.
First, its stabilizer generators in the bulk of the honeycomb lattice have weight six, compared to weight four in the surface code, which leads to a larger number of possible hook error combinations per stabilizer.
Second, equivalent logical operators in the color code can propagate across the system in several directions, whereas in the surface code they follow a path along a specific direction.

Asymptotically, the color code uses $\frac{3}{4}d^2$ data qubits to achieve a (phenomenological) code distance $d$, compared with $d^2$ for the rotated surface code.
A more informative figure of merit, however, is the total number of qubits $n_{\mathrm{tot}}$ (data and auxiliary) divided by the square of the circuit-level distance $d_{\mathrm{circ}}$ (the minimum number of circuit faults that can cause a logical error).
A hook error is an error on an auxiliary qubit midway through syndrome extraction that propagates through the remaining two-qubit gates into a correlated error on multiple data qubits.
In the surface code, a favorable syndrome extraction schedule can avoid all malign hook errors that lead to circuit-level distance reduction~\cite{kishony2025surface}.
In the color code, by contrast, any spatially uniform circuit that uses the conventional single auxiliary qubit per plaquette for syndrome extraction suffers a reduction in the circuit-level code distance by a factor of two, which undermines its competitive edge.

Much work has been done on the design of syndrome extraction circuits for the color code \cite{Baireuther_2019, beverland2021cost, lee2025colorcode, gidney2023colorcode, yoshida2025lowdepthcolorcode, PRXQuantum.5.030352}.
In particular, Beverland et al.~\cite{beverland2021cost} and Lee et al.~\cite{lee2025colorcode} optimized spatially uniform single-auxiliary circuits (three qubits per plaquette), achieving substantial improvements in logical error rate, yet the distance-halving limitation remains.
Gidney and Jones~\cite{gidney2023colorcode} proposed circuits that are not of this form: the middle-out'' circuit uses zero auxiliaries per plaquette (two qubits per plaquette), and the superdense'' circuit uses two auxiliaries per plaquette (four qubits per plaquette).
The middle-out circuit suffers the same halved distance but reduces the total number of qubits required to achieve a given circuit-level distance.
The superdense circuit achieves the full circuit-level distance by measuring both X and Z stabilizers on the same plaquette in parallel, with each auxiliary additionally serving as a flag for hook errors acting on the other; this comes at the cost of extra qubits.
Since the conventional and middle-out circuits have $d_{\mathrm{circ}} = d/2$, they achieve $n_{\mathrm{tot}}/d_{\mathrm{circ}}^2 \sim \frac{9}{2}$ and $n_{\mathrm{tot}}/d_{\mathrm{circ}}^2 \sim 3$, respectively; for the superdense circuit, $d_{\mathrm{circ}} = d$, giving $n_{\mathrm{tot}}/d_{\mathrm{circ}}^2 \sim \frac{3}{2}$.

In this work, we propose a syndrome extraction circuit that uses a single auxiliary qubit per plaquette yet maintains the full circuit-level distance in the bulk, achieving $n_{\mathrm{tot}}/d_{\mathrm{circ}}^2 \sim \frac{9}{8}$.
The crucial insight is that different hook errors are malign on plaquettes of different colors, and therefore different gate schedules should be used for plaquettes of each color.
We construct a circuit for which all hook errors in plaquettes of each color are benign.
We maintain the minimal circuit depth possible with a single ancilla per plaquette: all Pauli-X stabilizers are measured in parallel in six time steps, and all Z stabilizers are measured in parallel in six time steps.

\begin{figure*}
  \centering
  \begin{tikzpicture}
    \node[anchor=south west,inner sep=0] (img) at (0,0) {\includegraphics[width=\textwidth]{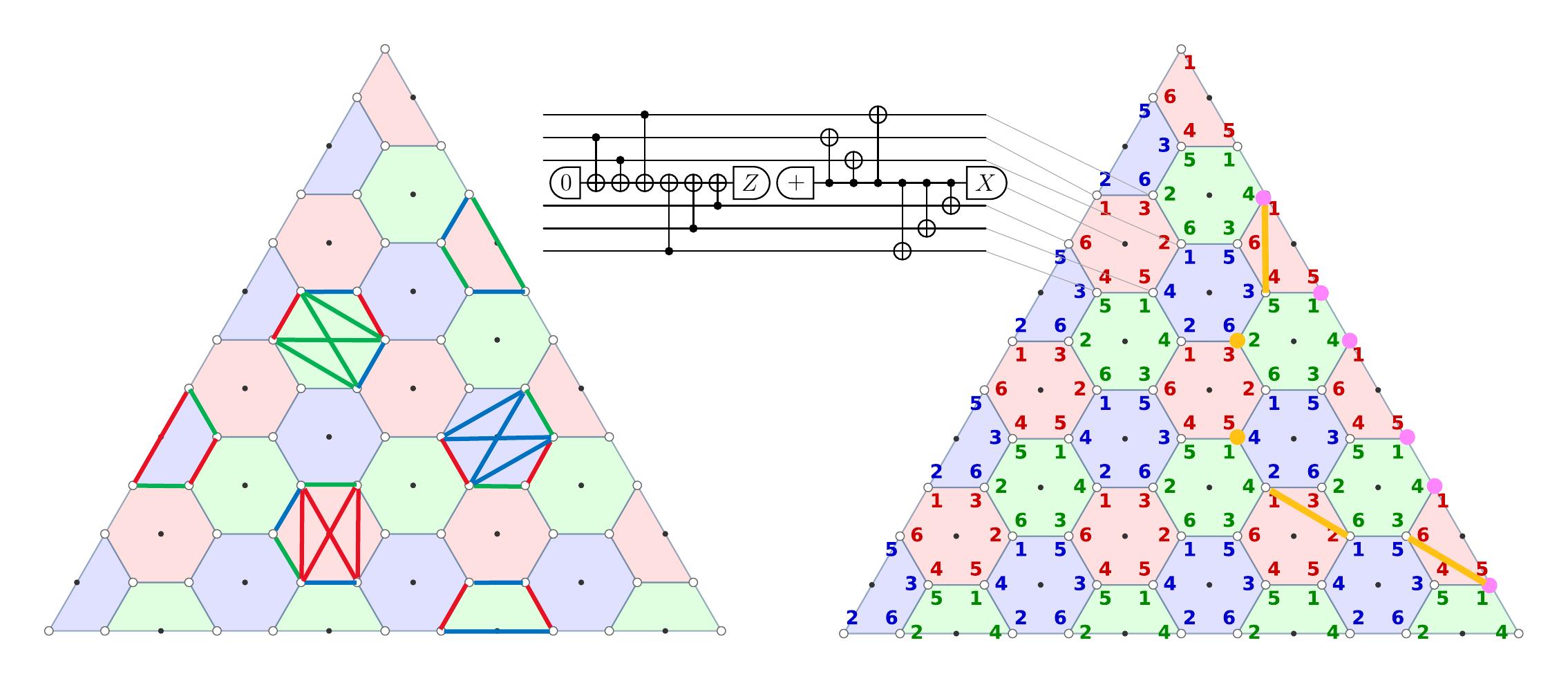}};
    \node[anchor=south west,font=\LARGE\bfseries] at ([xshift=0.15\textwidth,yshift=-30pt]img.north west) {a};
    \node[anchor=south west,font=\LARGE\bfseries] at ([xshift=0.67\textwidth,yshift=-30pt]img.north west) {b};
  \end{tikzpicture}
  \caption{\textbf{Malign hook errors and our schedule.}
  \textbf{(a)} Illustration of all malign hook error pairs in plaquettes of each color (triplets not shown as they are benign).
  Each highlighted pair would reduce the distance by one if it could be generated by a hook error on the plaquette hosting that pair.
  The pairs themselves are colored according to the type of anyon for which they would provide a shortcut.
  Malign hook errors on trapezoidal boundary plaquettes differ slightly from those in the bulk.
  On the corner plaquettes, all hook errors are malign (not shown).
  \textbf{(b)} Our syndrome extraction circuit.
  All malign hook errors are avoided in the bulk, and the depth is minimal.
  On the boundary certain combinations of hook errors can reduce the distance slightly; we call these ``fractional hook errors.''
  Three hook errors (orange lines) and two data-qubit errors (orange dots) are equivalent to a sequence of six data-qubit errors (purple dots) along the boundary, forming part of a logical operator.
  }
  \label{fig:malign_and_schedule}
  \end{figure*}



\begin{figure}[t]
  \centering
  \begin{tikzpicture}
    \node[anchor=south west,inner sep=0] (img) at (0,0) {\includegraphics[width=\columnwidth]{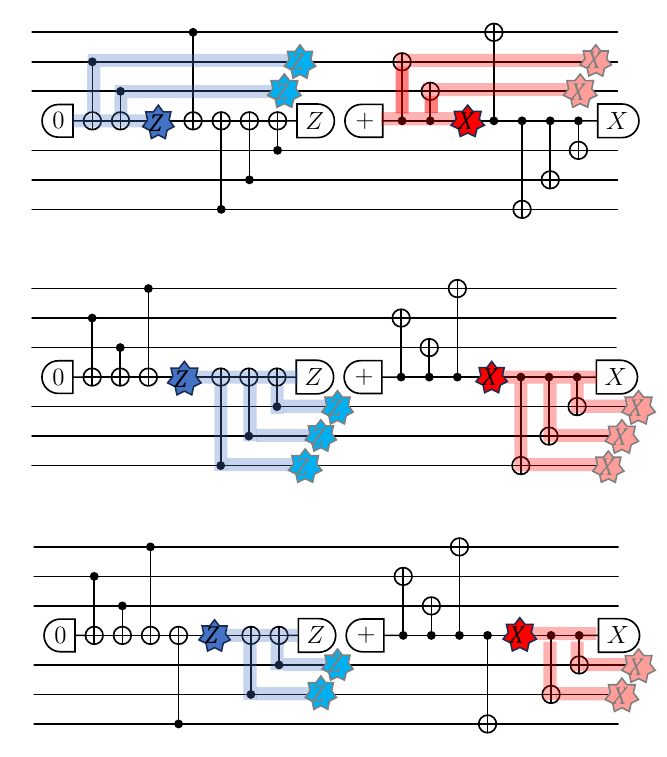}};
    \node[anchor=south west,font=\LARGE\bfseries] at ([xshift=0.06\columnwidth,yshift=-12pt]img.north west) {a};
    \node[anchor=south west,font=\LARGE\bfseries] at ([xshift=0.06\columnwidth,yshift=-105pt]img.north west) {b};
    \node[anchor=south west,font=\LARGE\bfseries] at ([xshift=0.06\columnwidth,yshift=-200pt]img.north west) {c};
  \end{tikzpicture}
  \caption{\textbf{Hook error propagation.}
  In each plaquette, there are three possible hook errors that can act on the auxiliary qubit during the measurement of each of the two Pauli stabilizers (dark blue/red).
  Each of these errors propagates through the syndrome extraction circuit to form a distinct correlated data-qubit error (light blue/red).
  \textbf{(a)} Errors occurring after the second CNOT during the measurement of each stabilizer.
  \textbf{(b)} Errors between the third and fourth CNOTs.
  \textbf{(c)} Errors between the fourth and fifth CNOTs.}
  \label{fig:all_hook_errors}
\end{figure}

\section{Malign hook errors and schedule design}
\label{sec:malign}

The 2D color code is defined on the honeycomb lattice [see Fig.~\ref{fig:malign_and_schedule}]: data qubits sit on the vertices, and the faces can be colored red, green, and blue such that no two adjacent faces share a color.
Likewise, each edge is assigned the common color of the two plaquettes it touches at its endpoints.
For each face $f$, the code has an X-type and a Z-type stabilizer generator, $S_f^X = \prod_{v \in f} X_v$ and $S_f^Z = \prod_{v \in f} Z_v$, so that each bulk stabilizer has weight six.
Since the color code is a Calderbank-Shor-Steane (CSS) code, we can consider the effects of Pauli X and Z errors separately.
Furthermore, because the code is self-dual, X and Z errors are equivalent.

A Pauli X error on a single data qubit flips the value of the Z-type stabilizers on the three plaquettes that share that data qubit.
These excitations are called red, green, and blue anyons according to the colors of the plaquettes on which they reside~\cite{Kesselring_2024}.
Anyons can be moved from one plaquette to another of the same color by Pauli X errors acting on data qubits at the ends of an edge connecting those plaquettes.
A single logical qubit is encoded on a triangular patch with three distinct boundary types—red, green, and blue—according to the color of the plaquettes missing from each boundary.
A logical X operator is a string-like product of Pauli X operators along a path of edges connecting the three boundaries; the code distance $d$ is the minimum weight of such an operator.
A logical operator can be viewed as propagating a red anyon from the red boundary, a green anyon from the green boundary, and a blue anyon from the blue boundary to the same point in the bulk, where all three annihilate together.
Equivalently, a logical operator can be written on the support of the data qubits along the entire length of one of the boundaries.

We consider syndrome extraction circuits of the conventional form: one auxiliary qubit per plaquette, where each stabilizer measurement is implemented by coupling that auxiliary to the six data qubits of the plaquette via six two-qubit gates and then measuring the auxiliary.
The same procedure is applied sequentially to X-type and Z-type stabilizers, as illustrated in Fig.~\ref{fig:malign_and_schedule}(a).
The order in which the six two-qubit gates are applied on each plaquette defines the syndrome extraction schedule for that plaquette.
A hook error is a Pauli error on the auxiliary partway through the six gates; it propagates through the gates applied after that point in the schedule, or equivalently through those applied before it, becoming a correlated error on the data qubits.
Because each stabilizer has weight six, a given schedule yields three possible hook error combinations, depending on whether the fault occurs after the second, third, or fourth gate; these result in correlated data-qubit errors of weight 2, 3, and 2, respectively, as illustrated in Fig.~\ref{fig:all_hook_errors}.

Our key observation is that different hook errors are malign (i.e. serve as a shortcut for a logical operator) on plaquettes of different colors.
We therefore propose using a different schedule for different plaquettes, in contrast to all previous work, which used a single schedule for all plaquettes.
Therefore we suggest using a {\it different schedule for different plaquettes}, in contrast to all previous work which used a single schedule for all plaquettes.
Since a low weight logical operator is supported on a direct path of edges connecting the three boundaries, its support includes at most two data qubits from any given plaquette.
For this reason, although the hook error triggered by a fault exactly halfway through the schedule of a plaquette produces the highest-weight correlated error, it is benign because it does not create a shortcut for a logical operator.
In other words, such an error typically flips a large number of neighboring stabilizers, which are then difficult to flip back with further errors.
Therefore, in our initial analysis below, we focus on the effects of weight-2 hook errors.

Fig.~\ref{fig:malign_and_schedule}(a) shows all malign weight-2 correlated error pairs in plaquettes of each color (in the bulk, we find that the set of malign pairs is the same for all plaquettes of a given color).
To understand why these errors are malign, first consider the highlighted pairs that coincide with edges of the lattice.
These pairs provide shortcuts for anyons propagating along those edges, that is, along edges of specific colors and orientations.
Shortcuts for logical operators are enabled by correlated errors on edges of each color that are oriented nearly perpendicular to the boundary of the same color.
Highlighted pairs that do not coincide with edges of the lattice can be viewed as two halves of edges leading outward from the plaquette hosting the correlated error.
Such a pair is malign when both outgoing half-edges, which share the color of the plaquette hosting the error, are oriented nearly perpendicular to the boundary of that color.

While each of these highlighted correlated errors individually reduces the circuit-level distance by one, we emphasize that most of them, if replicated across all plaquettes of the same color, would reduce the distance by a factor of two.
Furthermore, any weight-2 correlated error pattern is malign on plaquettes of some color.
This underlines the impossibility of avoiding the distance-halving problem with a single auxiliary qubit per plaquette using a {\it spatially uniform schedule}.

To preserve the full circuit-level distance in the bulk, we assign a different schedule to plaquettes of each color such that all malign hook errors are avoided [Fig.~\ref{fig:malign_and_schedule}(b)].
Furthermore, we design the circuit so that the schedules of neighboring plaquettes do not conflict, that is, they do not access the same data qubits at the same time, allowing all stabilizers of the same Pauli type to be measured in parallel in the minimal six two-qubit-gate steps.
On the boundaries of the system, some plaquettes are trapezoidal and have four data qubits rather than six.
For simplicity, we assign these plaquettes the same schedule as the bulk plaquettes of the same color, omitting the gates to the two absent data qubits.
Finally, we note that any weight-2 correlated error on these weight-4 plaquettes that coincides with one of their edges is malign.
We therefore choose the bulk schedule such that the corresponding schedule on the boundary plaquettes yields hook errors along their diagonals.
Subject to all of the constraints above, multiple valid schedules remain, of which we select one arbitrarily.
It may be interesting to explore the performance of the alternatives in future work.

While our schedule achieves the full circuit-level distance in the bulk, some hook errors at the boundaries can still reduce the distance slightly.
Because this reduction is confined to the boundary, its effect is expected to become negligible at large code distances.
Nonetheless, adding flag qubits on boundary plaquettes, at a subleading cost in qubit number, may be beneficial for finite system sizes and realistic noise models.
Interestingly, our circuit does not admit any single hook error that alone reduces the circuit-level distance, except at the corners of the system.
Instead, only certain combinations of several hook errors along the boundary can create a shortcut for a logical operator and thereby reduce the circuit-level distance.
We refer to these combinations as ``fractional hook errors.''
Specifically, we find that a combination of three weight-2 hook errors and two individual data-qubit errors near the boundary is equivalent to a sequence of six data-qubit errors along the boundary, forming part of a logical operator [see Fig.~\ref{fig:malign_and_schedule}(b)].
Therefore, we conclude that the circuit-level distance along the boundaries is asymptotically reduced by a factor of $5/6$.
Indeed, we verify numerically that, up to $d=13$, the circuit-level distance is given by $d_{\mathrm{circ}} = d - \left\lfloor \frac{d+3}{6} \right\rfloor$.

\begin{figure*}[t]
  \centering
  \includegraphics[width=\textwidth]{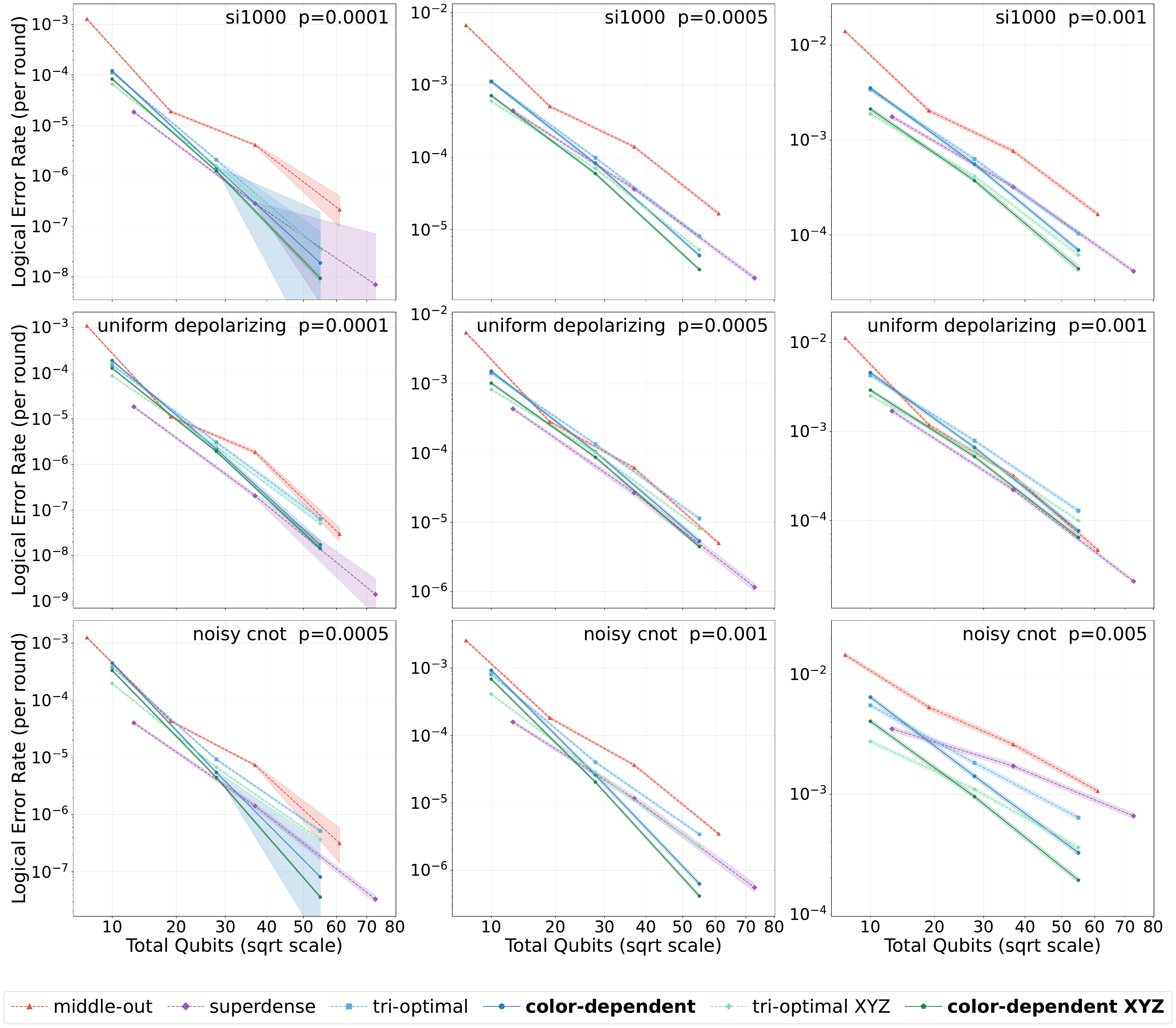}
  \caption{\textbf{Logical error rate vs.\ total qubit count across noise models and error rates.}
  Top row: SI1000 noise model. Middle row: uniform depolarizing noise model. Bottom row: noisy CNOT noise model.
  The physical error rate increases from left to right across the panels.
  Solid lines show our circuits, and dashed lines show previous work.
  Highlights indicate hypotheses with likelihoods within a factor of 1000 of the maximum-likelihood hypothesis, given the sampled data.}
  \label{fig:error_vs_qubits}
\end{figure*}

\section{Simulations}
\label{sec:simulations}

We compare our color-dependent circuit against the three other circuit constructions discussed above using Monte Carlo simulations: the optimized uniform single-auxiliary circuit~\cite{lee2025colorcode} (tri-optimal), the middle-out circuit~\cite{gidney2023colorcode}, and the superdense circuit~\cite{gidney2023colorcode}.
We also simulate the XYZ versions~\cite{gidney2023colorcode} of the tri-optimal circuit and the color-dependent circuit.
In these versions, instead of measuring X and Z stabilizers in each round, we cycle through measurements of the X, Y, and Z stabilizers, taking advantage of the self-dual nature of the color code.
The tri-optimal XYZ circuit was established as the state of the art for the realistic superconducting-inspired (SI1000) noise model~\cite{Gidney2021faulttolerant} in Ref.~\cite{koutsioumpas2025colourcodesreachsurface}.

We simulate logical idling (memory experiments) in the Z basis under circuit-level noise using Stim~\cite{gidney2021stim}.
We run $d$ rounds of syndrome extraction per shot and report logical error rates per round.
Decoding is performed with the Tesseract decoder~\cite{tesseract2024}, which can be readily applied to all six circuits for a fair comparison.
The decoder receives syndrome information from all measured stabilizers in all circuits; this is required for the flagging of hook errors in the superdense circuit.
The circuits and Python code used to generate these results are available at \url{https://github.com/Classiq/classiq-library/tree/main/algorithms/error_correction/}.

We test the performance of the circuits under three different circuit-level noise models: the SI1000 noise model, the uniform depolarizing noise model, and the noisy CNOT noise model [see Appendix~\ref{sec:noise_models}].
Fig.~\ref{fig:error_vs_qubits} shows the logical error rate per round versus the total number of qubits, with solid lines corresponding to our circuits and dashed lines to previous work.
The data are shown for several physical error rates for each noise model.

For all noise models and physical error rates considered, our circuits already match or outperform the previous state of the art at the modest code distances simulated here, in terms of the qubit footprint required to reach a given logical error rate.
Because the logical error rate decreases more favorably with qubit count for our circuits, as is evident from the data, this advantage should grow with increasing distance.
This behavior is consistent with the analytical argument above relating footprint to circuit-level distance, although that argument determines the scaling of logical error rate with distance only in the asymptotically low physical error rate regime.
Nevertheless, our circuits perform well even at relatively high physical error rates that are achievable with current hardware.
Since our circuit suffers reduced distance only at the boundaries of the system, its performance is expected to improve further at larger distances, where boundary effects are suppressed; the corresponding lines, which curve slightly downward with increasing qubit count, are consistent with this expectation.

The advantage of our circuits is most clearly visible in the noisy CNOT noise model.
While this model includes errors equivalent to faults at all circuit locations with probability proportional to $p$, the effect of hook errors is highlighted because they are no less likely than other faults, unlike in the SI1000 model, where measurement errors are relatively likely.
In the noisy CNOT model, the difference in scaling between our circuits and the alternatives is therefore most pronounced.
By contrast, in the uniform depolarizing noise model, the performance differences between the circuits are quite small.
For the realistic SI1000 noise model, which is of practical interest, we find by extrapolation to low logical error rates that the teraquop footprint achieved by our circuits is roughly 20\% smaller than that of the next-best circuit across the physical error rates considered.

\section{Summary}
\label{sec:summary}

We have presented a syndrome extraction circuit for the 2D color code that uses a single auxiliary qubit per plaquette yet maintains the full circuit-level distance in the bulk.
The key idea is to assign a distinct CNOT schedule to each plaquette color, chosen to avoid the hook errors that are malign for that color.
Our circuit has minimal depth: all stabilizers of the same Pauli type are measured in parallel in six time steps.
At the boundary, no single hook error reduces the distance; only combinations of hook errors, which we call fractional hook errors, do.
We have demonstrated through Monte Carlo simulations that our circuit outperforms the previous state-of-the-art circuits.
We also adapt the XYZ color code circuit to use our schedule, achieving a further reduction in footprint in some cases.

We note that, while in the surface code the effects of malign hook errors generated by a poor schedule can be mitigated by alternating between a schedule and its reverse~\cite{bluvstein2025architectural, gidney2024alternating}, the same trick does not work in the color code.
This is because, in the conventional syndrome extraction circuit for the color code, the X and Z stabilizers are measured in separate rounds, so hook errors are purely spatial rather than diagonal in space-time.

In future work, it may be interesting to study the performance of our circuit in various lattice-surgery procedures, as well as in magic state cultivation.
It may also be possible to combine the scheduling strategy from this work together with the structure of the circuits proposed in Ref.~\cite{gidney2023colorcode} which are designed with square-lattice connectivity in mind.
We hope that the insights from this work, together with a similar analysis of the nature of hook errors, can help guide the design of more efficient syndrome extraction circuits for other code families.
These ideas may also prove useful alongside automated tools for the design of syndrome extraction circuits, such as those in Refs.~\cite{prophunt2026,alphasyndrome2026}.

\acknowledgments

We thank Ron Cohen, Shoham Jacoby, Peter-Jan H. S. Derks, Stergios Koutsioumpas, and Craig Gidney for useful discussions.

\appendix

\section{Noise models}
\label{sec:noise_models}

We use three circuit-level noise models in our simulations. In each case, a single parameter $p$ sets the physical error rate.

\paragraph{SI1000 (superconducting-inspired).}
The SI1000 model~\cite{Gidney2021faulttolerant} is designed to capture typical noise in superconducting qubits.
Idle qubits undergo depolarization at rate $p/10$, with an additional depolarization of $2p$ when idling during a measurement or reset operation on other qubits.
Single-qubit Clifford gates are followed by single-qubit depolarizing noise with probability $p/10$, and two-qubit gates are followed by two-qubit depolarizing noise with probability $p$.
Measurement outcomes are flipped with probability $5p$.
Reset operations prepare a flipped state with probability $2p$.

\paragraph{Uniform depolarizing.}
The uniform depolarizing model assigns the same error probability $p$ to every circuit location: idle depolarization, single-qubit depolarization after each one-qubit gate, two-qubit depolarization after each two-qubit gate, flipped measurement outcomes, and reset errors.

\paragraph{Noisy CNOT.}
The noisy-CNOT model applies two-qubit depolarizing noise with probability $p$ after every CNOT gate only.
While this model is minimalistic, it still generates all circuit-level faults captured by the other two models with probability linear in $p$.

\bibliography{references}

\end{document}